\documentclass[10pt,onecolumn,superscriptaddress,amsmath,amssymb,enumerate]{revtex4}
\usepackage{graphicx}
\usepackage{dcolumn}
\usepackage{bm}
\usepackage{color}
\usepackage{url}
\usepackage{tabularx}
\usepackage{array}
\usepackage{booktabs}
\usepackage{comment}
\usepackage{braket}
\usepackage{amsmath}
\usepackage[colorlinks=true, linkcolor=blue, citecolor=blue, urlcolor=blue]{hyperref}

\newcommand{\vcc}[1]{\mathbf{#1}} 			
\newcommand{\tdd}[2]{\frac{d #1}{d #2}}	
\newcommand{\Ham}{\mathcal{H}}		    
\date{\today}	

\begin{document}

\title{Switching of chiral magnetic skyrmions by picosecond magnetic field pulses via transient topological states}

\author{Changhoon Heo}
\thanks{These authors contributed equally to this work}
\affiliation{Radboud University, Institute for Molecules and Materials, Heyendaalseweg 135, 6525 AJ, Nijmegen, The Netherlands}

\author{Nikolai S. Kiselev}
\thanks{These authors contributed equally to this work}
\affiliation{Peter Gr\"unberg Institut and  Institute for Advanced Simulation, Forschungszentrum J\"ulich and JARA, D-52425
  J\"ulich, Germany}
  
\author{Ashis Kumar Nandy}
\thanks{These authors contributed equally to this work}
\affiliation{Peter Gr\"unberg Institut and  Institute for Advanced Simulation, Forschungszentrum J\"ulich and JARA, D-52425
  J\"ulich, Germany}

\author{Stefan Bl\"ugel} 
\affiliation{Peter Gr\"unberg Institut and  Institute for Advanced Simulation,  Forschungszentrum J\"ulich and JARA, D-52425
  J\"ulich, Germany}
 
\author{Theo Rasing} 
\affiliation{Radboud University, Institute for Molecules and Materials, Heyendaalseweg 135, 6525 AJ, Nijmegen, The Netherlands}

\begin{abstract}

\textbf{Magnetic chiral skyrmions are vortex like spin structures that appear as stable or meta-stable states in magnetic materials due to the interplay between the symmetric and antisymmetric exchange interactions, applied magnetic field and/or uniaxial anisotropy. Their small size and internal stability make them prospective objects for data storage but for this, the controlled switching between skyrmion states of opposite polarity and topological charge is essential. Here we present a study of magnetic skyrmion switching by an applied magnetic field pulse based on a discrete model of classical spins and atomistic spin dynamics. We found a finite range of coupling parameters corresponding to the coexistence of two degenerate isolated skyrmions characterized by mutually inverted spin structures with opposite polarity and topological charge. We demonstrate how for a wide range of material parameters a short inclined magnetic field pulse can initiate the reliable switching between these states at GHz rates.  Detailed analysis of the switching mechanism revealed the complex path of the system accompanied with the excitation of a chiral-achiral meron pair and the formation of an achiral skyrmion.}

\end{abstract}
  
\maketitle

Magnetic chiral skyrmions are vortex like spin textures with particle like properties. They may appear as stable (hexagonal lattice of skyrmions) or metastable states (isolated skyrmions) in magnetic materials as a result of the interplay between  Heisenberg exchange, Dzyaloshinskii-Moriya interaction (DMI) \cite{Dzyaloshinskii,Moriya}, applied magnetic field and/or uniaxial anisotropy. Their nontrivial topology gives rise to intriguing dynamic properties such as topological Hall effect \cite{TopHall}, skyrmion Hall effect \cite{SkHall} and unconventional electromagnetic behavior \cite{Jonietz_10, Schulz_12}. The theory of thermodynamic stability of magnetic skyrmions was developed by Bogdanov and coworkers \cite{Bogdanov_89,Bogdanov_94,Bogdanov_99,Bogdanov_06, Bogdanov_11, Bogdanov_14}. The interest in skyrmion properties and potential applications strongly increased by several indirect \cite{Muhlbauer_09} and direct \cite{Yu_10,Yu_11,Romming_13} observations of skyrmions with different techniques in different materials. 
Magnetic skyrmions are attractive for use in spintronic devices because of their high mobility for low current densities \cite{Yu_12} and \textit{internal} stability \cite{Kiselev_11}, 
as the relevant	interactions, in general, do not depend on the size and shape of the sample.
Recently, Fert and coauthors presented a conceptual idea of a spintronic device based on skyrmion motion driven by spin-polarized currents \cite{Fert_13}, similar to the racetrack memory based on the domain wall motion \cite{Parkin_15} but much more energy efficient.
Skyrmions in such a device are assumed to be stabilized on top of a ferromagnetic ground state of fixed magnetic polarization.

Here, we propose an alternative scheme which involves the manipulation of an isolated skyrmion (iSk) as a data bit localized in a finite-size domain, similar to an element of Magnetoresistive Random Access Memory (MRAM) \cite{MRAM}.
We demonstrate the stability of such iSk in zero magnetic field and the possibility of switching between two degenerate skyrmion states characterized by opposite polarity and topological charge ($Q$) by an inclined magnetic pulse. 
We found that the switching between such states takes place via the excitation of a pair of chiral and achiral merons and the subsequent emergence of a transient intermediate achiral skyrmion. 
Our findings not only explain the microscopic details of the controlled switching process which can be achieved at GHz frequencies but also indicate the possibility of creating an MRAM type of device based on the manipulation of such skyrmion states without the necessity to apply a stabilizing magnetic field.

\begin{figure*}[ht]
 \centering
  \includegraphics[width=17.5cm]{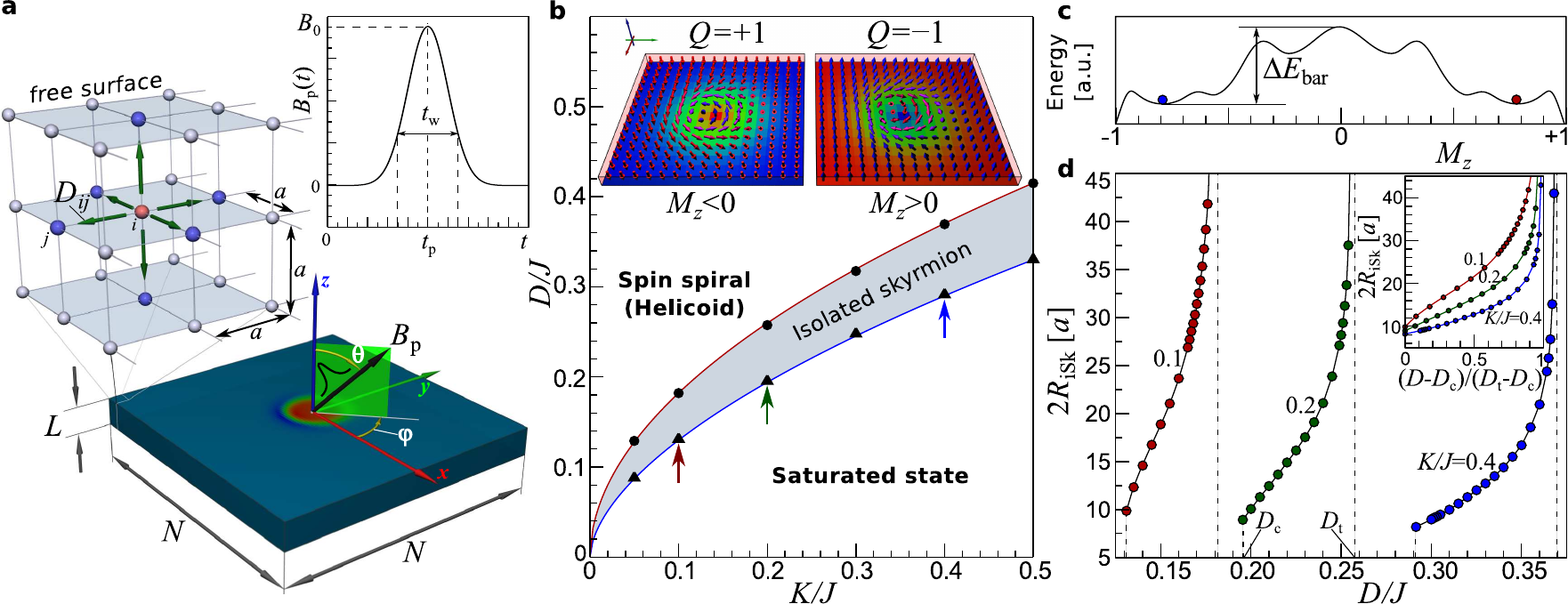}
 \caption{\textbf{Simulation schemes and stability of chiral magnetic skyrmions at zero magnetic field: }
(a) Schematic representation of the simulated system. The domain is composed of $N \times N$ atomic sites along the $x$ and $y$ -axes and 3 atomic monolayers thick with a simple cubic lattice of lattice constant $a$.
Each internal magnetic atom (see {\it e.g.} red sphere) has six nearest neighbors (blue spheres), while atoms at the surface and the edges have less neighbors.
Dzyaloshinskii-Moriya vectors (green arrows) point along the directions of nearest neighbors.
The magnetic filed pulse $\mathbf{B}_\mathrm{p}$ penetrates uniformly through the whole domain.
The direction of the pulse is defined by the polar angle $\theta$ and azimuthal angle $\varphi$.
The inset illustrates the Gaussian pulse with width $t_\mathrm{w}$, time of the pulse maximum $t_\mathrm{p}$, and amplitude $B_\textrm{0}$.
(b) The phase diagram of the ground state for zero applied field calculated in reduced units of DMI and uniaxial anisotropy for a three layer thick film. The red line corresponds to the second order phase transition between spin-spiral and saturated ferromagnetic state.
The blue line corresponds to the collapse for iSk.
The shaded area corresponds to the range of stability for iSk.
The inset shows two types of iSks solutions characterized by mutually inverted spin structures with opposite out-of-plane magnetization and topological charge. The red (blue) color of the background area denotes the positive (negative) component of the magnetization $m^\textrm{z}_i$.
(c) Schematic representation of the energy profile with many local minima. The ground state of the system corresponds to a two-fold degenerate saturated ferromagnetic state with $M_\textrm{z} = \pm 1$. Metastable solutions for iSks have equal energies and are separated by a finite energy barrier $\Delta E_\textrm{bar}$.
(d) Diameter of iSk, $2R_\textrm{iSk}$, as a function of reduced DMI constant $D/J$ for different values of reduced anisotropy $K/J=0.1$, $0.2$, $0.4$ (see arrows in Fig.~\ref{fig_stability}b).
The definition for iSk size on a discrete lattice is presented in Supplementary Materials \textbf{S2}.}
\label{fig_stability}
\end{figure*}

In our model, we consider a thin film of a chiral magnet, assuming a simple cubic lattice structure with lattice constant $a$, see Fig. 1a.
Direct energy minimization of the model Hamiltonian presented in \textbf{Method} allows one to identify the ground state of the system with respect to geometrical parameters, \textit{e.g.} thickness of the magnetic film, and material parameters as coupling constants of Heisenberg exchange ($J$), DMI ($D$) and uniaxial anisotropy constant ($K$).
Fig.~\ref{fig_stability}b shows the numerically calculated phase diagram of the ground state for an infinite magnetic film composed of three monolayers (thickness, $L = 2a$) at zero applied magnetic field, presented in terms of two reduced parameters $D/J$ and $K/J$. 
The red line corresponds to the second order phase transition between a spin spiral (SS) and a saturated ferromagnetic (FM) state. 
The period of the SS state goes to infinity and effectively approaches the FM state when $D/J$ or $K/J$ approaches the phase transition line.
In the case of a magnetic film of a finite thickness, the free surfaces where the magnetic atoms have a reduced number of neighbors provide an additional spatial modulation of the magnetization also through the whole film thickness \cite{Rybakov_13,Rybakov_15}. 
Due to the complexity of such a three-dimensional state only numerical calculations allow an identification of the correct phase transitions.
The details of the calculation of the phase diagram and a comparison with the analytical solution for the limiting cases of pure two-dimensional and bulk chiral magnets are given in Supplementary Materials \textbf{S1}.
According to the micromagnetic continuum model, a solution for metastable iSks can be found for any $D/J$ and $K/J$ below the phase transition line within the FM ground state \cite{Bogdanov_94,Bogdanov_99}.
However, the characteristic sizes of such skyrmion solutions can be smaller than the lattice constant and, thereby, lose their physical meaning.
In order to describe the stability of magnetic skyrmions properly, it is therefore crucial to use an adequate discrete model, which allows one to identify the collapse (blue) line in Fig.~\ref{fig_stability}b.
Therefore, iSks are stable only within a finite range of $D/J$ and $K/J$, given by the shaded area in the phase diagram.
Due to the absence of a reference magnetic field, the two iSks with mutually inverted spin structures with opposite sign of out-of-plane magnetization (polarity) and topological charge $Q$ are degenerate, see inset in Fig.~\ref{fig_stability}b. 
These metastable skyrmion states are separated by a finite energy barrier $\Delta E_\textrm{bar}$ defined by a priori not known complex energy landscape, Fig.~\ref{fig_stability}c, which strongly depends on the material parameters and geometry of the system.
In Fig.~\ref{fig_stability}d, the dependence of the size of an iSk is shown for fixed values of $K/J$ marked by arrows in the phase diagram.
The size of an iSk goes to infinity when the ratio $D/J$  approaches the transition line, $D_\textrm{t}$ and becomes very small close to the collapse line, $D_\textrm{c}$. 
For very small values of $K/J$ and $D/J$, the size of the iSk also significantly increases, see inset in Fig.~\ref{fig_stability}d, which requires gigantic size of a simulated domain \cite{Rybakov_13, Rybakov_15}.
To investigate the generic features of switching without loss of generality we adjusted the material parameters such  that the atomistic spin-dynamics simulations can be performed on a reasonably large domain of 100$\times$100$\times$3 spins. 

\begin{figure*}[ht]
 \centering
  \includegraphics[width=17.5cm]{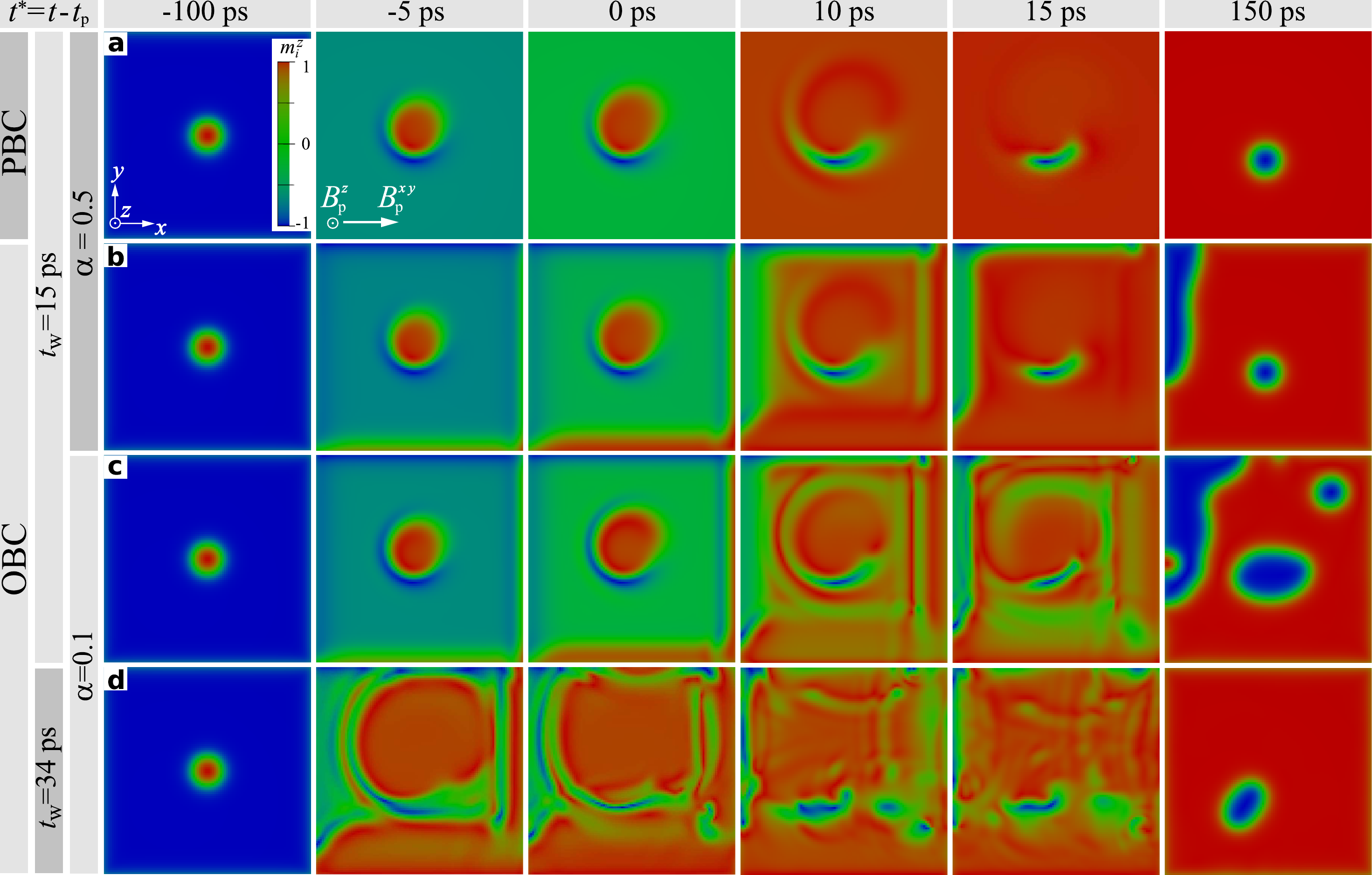}
  \caption{
  \textbf{Snapshots of the system during the dynamical switching induced by a single magnetic field pulse:}
The color represents the out-of-plane component of the magnetization, $m^\textrm{z}_i$ in the top atomic layer at each lattice site for the simulated domain of 100$\times$100$\times$3 spins.
The coordinate system and the color map for $m^\textrm{z}_i$ are shown in the initial snapshot of panel a.
The sequence of the snapshots in (a) corresponds to periodic boundary conditions (PBC) in the $xy$-plane, while in (b)-(d) open boundary conditions (OBC) are used.
The magnetic field pulse of amplitude $B_\textrm{0}=1$ T is applied along the direction defined by $\theta = 50^{\circ}$ and $\varphi = 0^{\circ}$.
The pulse width $t_{w}$ = 15 ps, and 34 ps, damping parameter $\alpha$ = 0.5 and 0.1 as marked on the left panel of the figure, see also open circles in Fig.~\ref{fig_switch_diagram_alpha}. For these simulations the coupling constants have been fixed to $J = 0.5$ meV/atom, $K/J = 0.2$, $D/J = 0.24$.}
 \label{fig_snapshots}
\end{figure*}

We have used a time dependent Gaussian magnetic field pulse $\textbf{B}_\textrm{p}$ applied in a direction defined by the polar angle $\theta$ and the azimuthal angle $\varphi$, as shown in Fig.~\ref{fig_stability}a, see also \textbf{Method}.

Fig.~\ref{fig_snapshots} shows the time-dependent snapshots of the switching process for different pulse widths $t_\textrm{w}$, damping parameter $\alpha$ and different boundary conditions: open (OBC) or periodic (PBC), see left panel in Fig.~\ref{fig_snapshots}.
Here and below, the origin of time is given relative to the time of maximum field pulse, $t^\ast=t-t_\textrm{p}$.
The initial (left) image shows an iSk with the core magnetization pointing up (red color) while the host is a ferromagnetic state with magnetization 
pointing down (blue color). 
The final state shown at the right most image is an almost relaxed state at 150 ps after the pulse maximum. 
The initial and final states represent mutually inverted spin structures, see inset Fig.~\ref{fig_stability}b.

Following the case of PBC, see the sequence in Fig.~\ref{fig_snapshots}a, we explain the main mechanism of the skyrmion switching which is based on two simultaneous processes: i) an expansion of the skyrmion core and ii) a homogeneous rotation of magnetization of the surrounding ferromagnetic state.
Due to the inclination of the applied field, the core of the skyrmion expands asymmetrically along the direction perpendicular to the projection of the magnetic field on the film surface, $B^{xy}_\textrm{p}$. 
Opposite to that direction one observes the formation of a region with magnetization opposite to the expanded core, see the blue area at $t^\ast$=-5 and 0 ps. 
Such an excited state has similarity to the vortex-antivortex pair, which can be observed in magnetic vortex core reversal dynamics \cite{Waeyenberge_06}.
Subsequently, the magnetization of the surroundings keeps on turning towards the field direction together with an expansion of the core.
When the surrounding magnetization is completely flipped, only the small blue region remains with opposite polarity.
Below, we show that this crescent shape object in the snapshots at $t^\ast$ = 10 and 15 ps corresponds to a non-axisymmetric achiral skyrmion which during the relaxation converges to an axisymmetric chiral skyrmion, see $t^\ast$ = 150 ps, with $Q$ and polarity opposite to the initial state.   

In the case of OBC, Fig.~\ref{fig_snapshots}b, the switching mechanism remains qualitatively the same as for the PBC, but the change in the polarity of the surrounding magnetization occurs in an inhomogeneous manner. First, one observes the appearance of an up-polarized magnetic region at the free boundary, see the red area at the bottom edge in Fig.~\ref{fig_snapshots}b, $t^\ast$ = 0 ps. 
It expands and quickly propagates through the whole domain, which results in the switching of the surrounding magnetization, see $t^\ast=10$ and 15 ps. 
Finally, the spin structure relaxes to an iSk and an additional domain may attach to the edge, see the final image in Fig.~\ref{fig_snapshots}b.
Nevertheless, within a short time of about 100 ps, this domain disappears and only a single chiral skyrmion remains.

In case of realistic damping, $\alpha \leq$ 0.3, a strong effect of spin-wave interference is observed, see Fig.~\ref{fig_snapshots}c and d. 
The spin waves injected and reflected by the free edges of the structure interfere with each other and interact with the expanded core of the skyrmion. 
This may lead to the skyrmion collapse and/or nucleation of new skyrmions. 
The final state may appear as a \textit{mixed} state, composed of multiple skyrmions and domain walls, see \textit{e.g.} the final image in Fig.~\ref{fig_snapshots}c. 
However, a successful \textit{one-to-one} switching can be controlled, for example by adjusting the parameters of the magnetic field pulse. 
Figure~\ref{fig_snapshots}d shows how, by adjusting the pulse width, the multiple skyrmion formation is suppressed and a one-to-one switching is observed.

\begin{figure*}[!t]
 \centering
  \includegraphics[width=17.5cm]{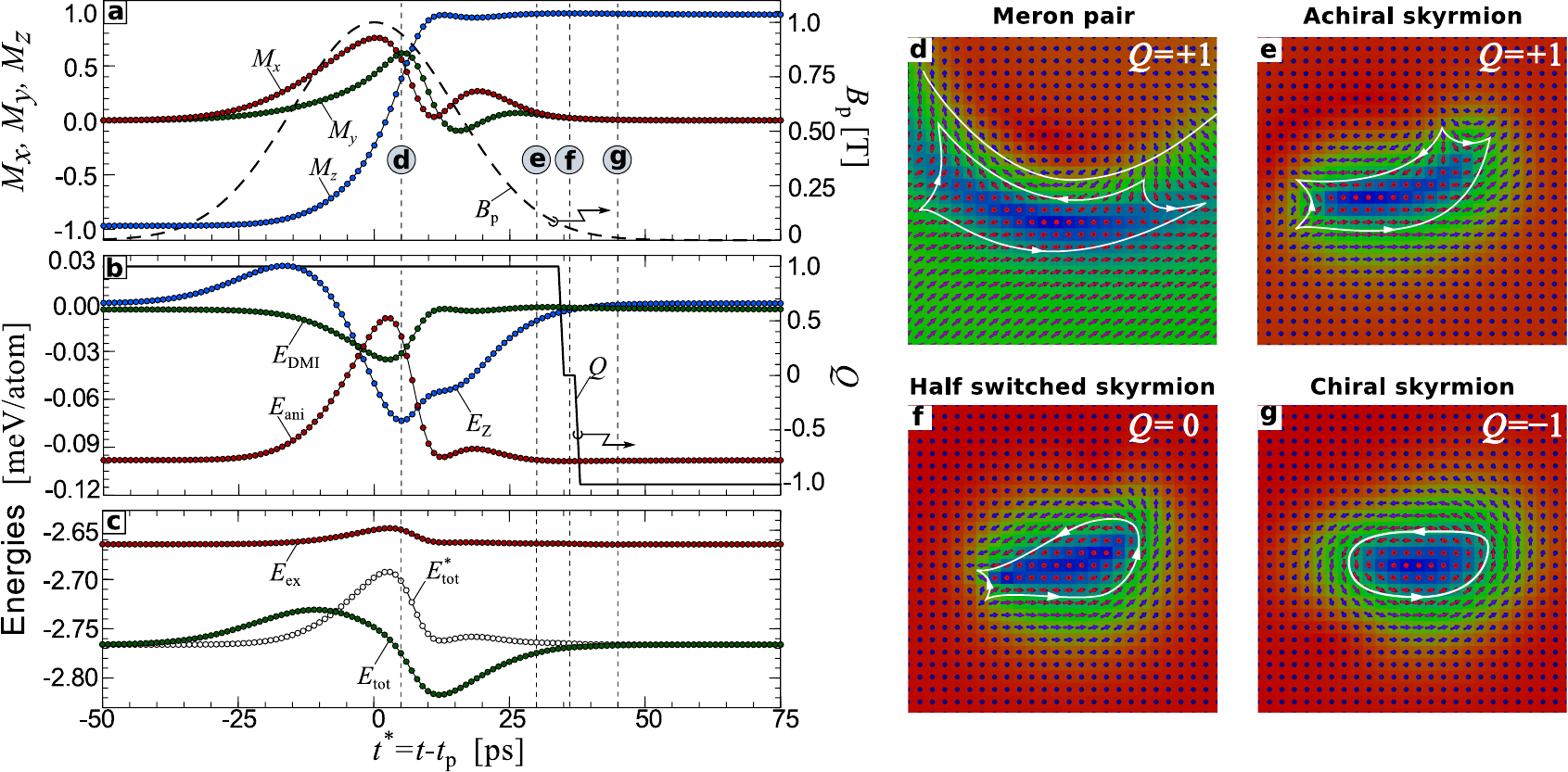}
 \caption{
 \textbf{Time dependencies of energy contributions and topological charge during the skyrmion switching:} (a)-(c) Time dependencies corresponding to the simulation of skyrmion switching with parameters as in Fig.~\ref{fig_snapshots}a.
Here $M_\textrm{x}$, $M_\textrm{y}$, $M_\textrm{z}$ are components of the average magnetization, $M_{\alpha}=\sum^N_i m^{\alpha}_{i}/\mu_sN$, $N$ total number of spins, $B_\textrm{p}$ is the profile of a magnetic field pulse,
and $Q$ is the topological charge.
$E_\textrm{DMI}$, $E_\textrm{ani}$, $E_\textrm{ex}$, and $E_\textrm{Z}$ correspond to the energy contributions of DMI, anisotropy, exchange, and Zeeman energy, respectively.
$E_\textrm{tot}$ is a sum of all energy contributions, while  $E^*_\textrm{tot}=E_\textrm{tot}-E_\textrm{Z}$.
(d)-(g) are the snapshots of the central part of the simulated domain (top atomic layer), taken during the simulation at the time marked in (a)-(c) by vertical dotted lines d-g, respectively.
\label{fig_time}
}
\end{figure*}

Reverse switching of skyrmions can be controlled by flipping the polar angle of the applied pulse with respect to the layer plane: $\theta^\prime=\pi - \theta$. 
With that, one can achieve skyrmion switching rates of the order of a few GHz. 
For instance, in our simulations the time interval between two field pulses required for sequential switching has typically a value of $200$ ps, which corresponds to a switching frequency of 5 GHz. 
In the Supplementary movies, we illustrate the back and forth skyrmion switching with an interval of $200$ ps driven by a sequence of pulses with alternating polar angles $\theta=45^\circ$ and $\theta^\prime=135^\circ$.

Such one-to-one skyrmion switching is by no means trivial: the two iSk states are metastable and the system switches between them across a complex energy landscape with various local minima and two global minima corresponding to the up and down ferromagnetic states.  
In order to understand better the general switching mechanism, we have studied the time dependence of the energy balance and the topological charge during the switching.  

\begin{figure*}[!t]
 \centering
  \includegraphics[width=17.0cm]{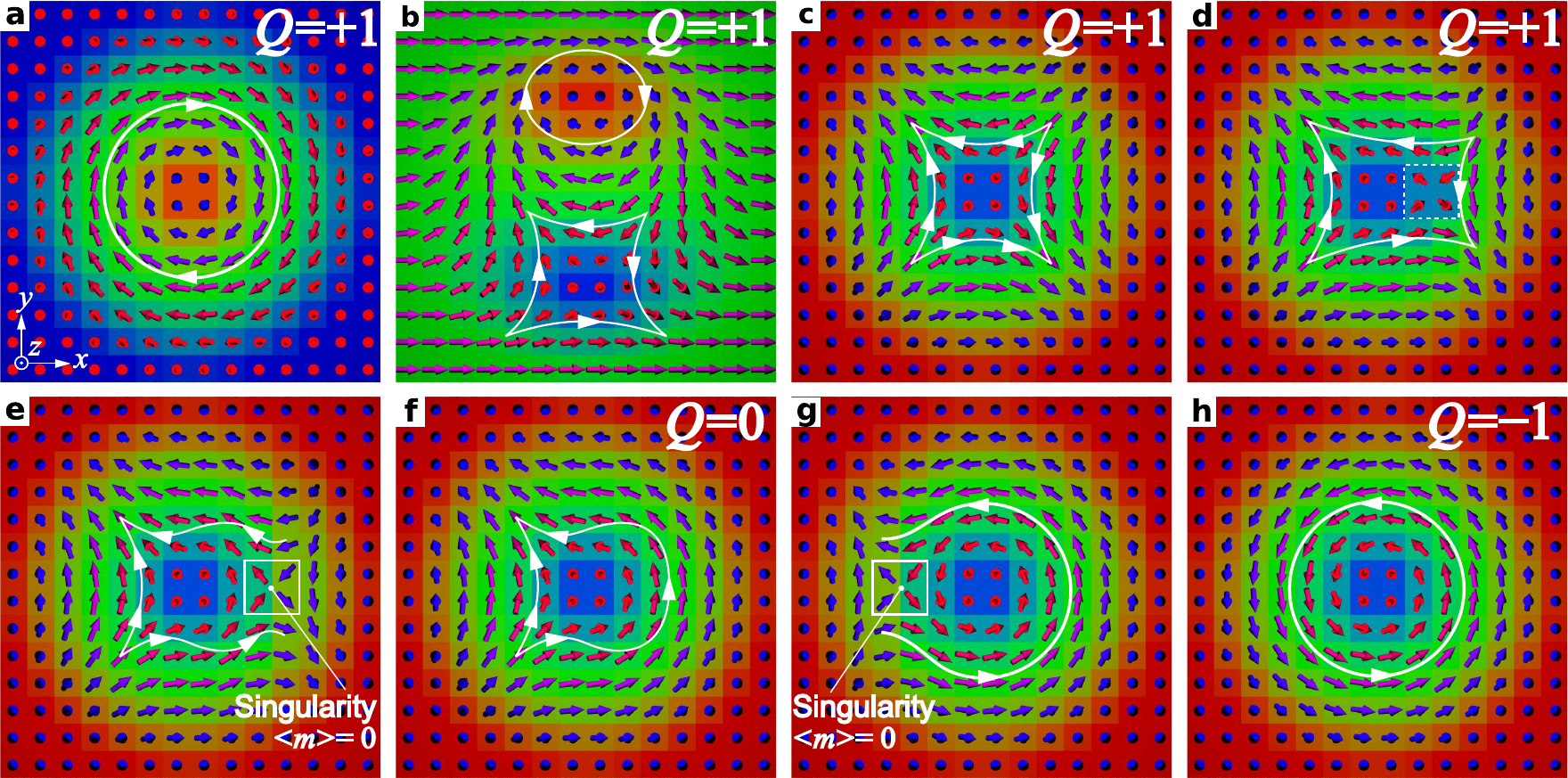}
 \caption{
 \textbf{Schematic representation of the transient topological states during the skyrmion switching:}
(a) and (h) are initial and final states of mutually inverted spin configurations corresponding to the chiral skyrmions with conserved chirality and opposite topological charge.
(b) - (g) the transient magnetic states during the switching:
(b) the chiral-achiral meron pair with total $Q$ = +1 obtained by 90$^\circ$ rotation of all spins in (a) around the $y$-axis;
(c) the achiral skyrmion with $Q$ = +1 obtained by continuing the rotation of all spins in (b) by another 90$^\circ$ around the $y$-axis;
(d) the achiral skyrmion state with local violation in the chirality inside the dashed box;
(e) the magnetic state with a singular point (SP), which makes the state topologically undefined;
(f) the half-switched skyrmion state with $Q$ = 0;
(g) the magnetic state with another SP appearing during the local change in the chirality of the spin structure.
The average magnetization in the white box is zero, $\braket{m}$ = 0.
The white isoline with arrows is the guide to the eye for the in-plane magnetization rotation direction.
\label{fig_4}
}
\end{figure*}

Figures~\ref{fig_time}a-c show the time dependencies for the magnetization, energy contributions, $Q$, and magnetic pulse profile corresponding to the simulation presented in Fig.~\ref{fig_snapshots}a.
The snapshots in Figs.~\ref{fig_time}d-g represent the central part of the spin structure at particular moments in time which are marked as vertical dashed lines in a-c.
The time dependence of $Q$ in Fig.~\ref{fig_time}b shows two steps at about $t^\ast=34$ ps and 38 ps, where $Q$ rapidly changes from $+1$ to $0$ and then from $0$ to $-1$.
The snapshots in Fig.~\ref{fig_time}e-g illustrate the changes in the spin texture before and after these steps.
To explain and clarify this step-like behavior, we schematically show the intermediate topological states through which the system passes during the switching in Fig.~\ref{fig_4}.
Figure~\ref{fig_4}a corresponds to the initial state and Fig.~\ref{fig_4}b represents a pair of \textit{chiral} and \textit{achiral} merons (or half skyrmions) carrying an individual topological charge $Q$ = $+1/2$, while the total $Q$ of the pair remains $Q$ = $+1$.
The schematic picture of this meron pair is structurally equivalent to the state in Fig.~\ref{fig_time}d (see also Fig.~\ref{fig_snapshots}a at $t^\ast = 0$ ps).
Because of the interaction with the inclined magnetic field, in the dynamical process the core of the chiral meron is much larger than the achiral one and both have distorted shapes. 
The state in Fig.~\ref{fig_4}c represents an ideal achiral skyrmion. 
The snapshot of an achiral skyrmion in our simulation is shown in Fig.~\ref{fig_time}e (see also Fig.~\ref{fig_snapshots}a at $t^\ast = 15$ ps). 	
\textit{The emergence of an achiral skyrmion via excitation of a meron pair is the key stage of the skyrmion switching mechanism}.
It reflects the tendency of the system to conserve $Q$, which in turn emphasizes the dominance of the Heisenberg exchange interactions. 
The initial chiral skyrmion, meron pair and achiral skyrmion belong to the same homotopy class, all three states possess the same $Q$. 
Due to the DMI, which provides the largest energy gain for those localized states with a single chirality, achiral skyrmion is energetically unfavorable and within a short time it switches to a chiral one.
Indeed, for an ideal achiral skyrmion as in Fig.~\ref{fig_4}c, the energy contribution of the DMI equals precisely zero, while for mutually inverted chiral skyrmions as in Fig.~\ref{fig_4}a and Fig.~\ref{fig_4}h, the energy gain from DMI is the same because of the conserved chirality of the spin structures.

An achiral skyrmion with $Q$=1 and a chiral skyrmion with $Q$=-1 belong to different homotopy classes and the transition between them may occur only via formation of a singular point (SP), where magnetization locally vanishes and $Q$ becomes undefined.
A local violation in the chirality of the spin structure, see dashed square in Fig.~\ref{fig_time}d, precedes the formation of a SP shown in Fig.~\ref{fig_time}e. 
The position of the SP can be associated with the center of a finite size volume where the average magnetization totally vanishes, see solid square in Fig.~\ref{fig_time}e.
Then, SP pushed out from the skyrmion into the surrounding ferromagnetic phase where where it smoothed out and disappear.
This, in turn, results in the formation of a so-called \textit{half-switched} skyrmion state with $Q$=0, see Fig.~\ref{fig_4}f and the corresponding simulation snapshot in Fig.~\ref{fig_time}f. The system remains in such intermediate state only for a short time (a few picoseconds, see $Q=0$ in Fig.~\ref{fig_time}b). 
The change in chirality on the other side of the half-switched skyrmion is also accompanied with the appearance of a SP, see Fig.~\ref{fig_4}g, and finally results in a transition to the chiral skyrmion, see Fig.~\ref{fig_4}h and snapshot in Fig.~\ref{fig_time}g. 
Such a transition results in the second jump of $Q$, from $Q=0$ to $-1$. 
In Fig.~\ref{fig_time}g, we have shown the chiral skyrmion just after the switching, which converges to an ideal axisymmetric skyrmion after a full relaxation.
	
The sequence of the states sketched above explains all the details of the evolution of the energy and magnetization in Fig.~\ref{fig_time}a-c. 
For instance, the excitation of a meron pair with a large number of spins pointing in the plane rather than in the easy out-of-plane direction results in a pronounced increase of the anisotropy energy $E_{\mathrm{ani}}$. 
Consequently, the chiral meron core expands which extends the inhomogeneous area and leads to a substantial lowering of the DMI energy $E_{\mathrm{DMI}}$ as well as to an increase in the Heisenberg exchange energy $E_{\mathrm{ex}}$. Subsequently, the formation of an achiral skyrmion and its localization leads to a lowering in $E_{\mathrm{ani}}$ and $E_{\mathrm{ex}}$ but causes an increase in the $E_{\mathrm{DMI}}$. 
The following transition between achiral and chiral skyrmions involves only a small number of spins around the SP and does not affect the total energy significantly.

\begin{figure}[!t]
 \centering
  \includegraphics[width=9.5cm]{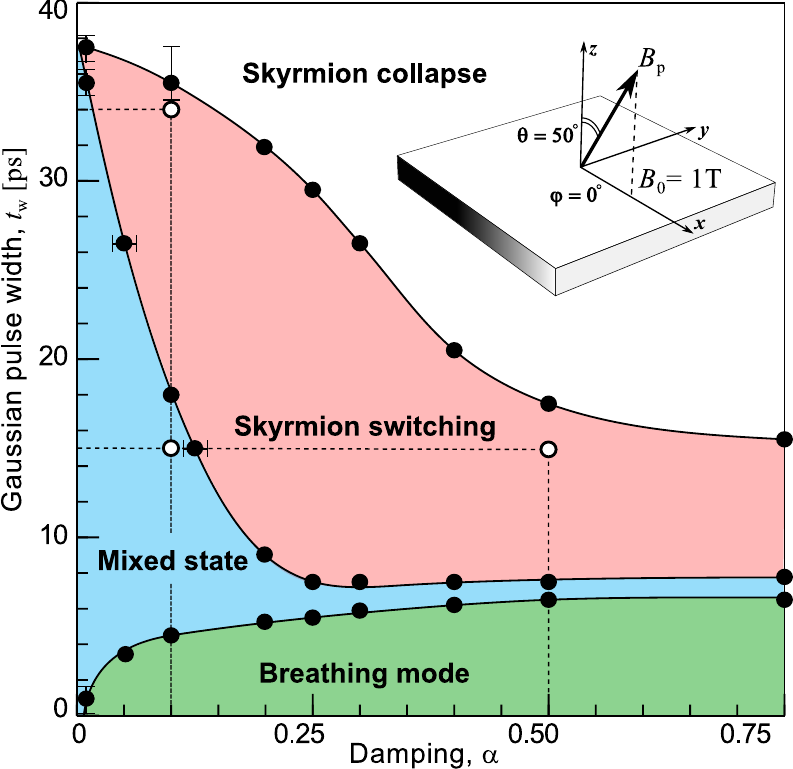}
 \caption{\textbf{Switching diagram for square domain in terms of damping $\alpha$ and magnetic pulse width $t_\textrm{w}$:} Each area corresponds to the regime of excitation: breathing mode (green), skyrmion switching (red), mixed state (blue) and skyrmion collapse (white).
Open circles correspond to the parameters used for the simulations illustrated by the snapshots in Fig.~\ref{fig_snapshots}.}
 \label{fig_switch_diagram_alpha}
\end{figure}

The precession of the spins around the applied magnetic field leads to additional excitations that affect the temporal behavior of the magnetization and energies. Such excitations become more pronounced in the case of OBC and show longer attenuation times for small damping $\alpha$. 
The influence of damping $\alpha$ and the pulse width $t_\textrm{w}$ on the induced dynamics is illustrated in the switching diagram presented in Fig.~\ref{fig_switch_diagram_alpha}.  
It shows four noticeable regions corresponding to the excitation of (i) the breathing mode, (ii) the \textit{mixed} state, (iii) the one-to-one skyrmion switching and (iv) the skyrmion collapse.
For short pulses, $t_\textrm{w}\lesssim 6$ ps, a skyrmion experiences only the breathing mode, \textit{i.e.} the excitation causes a skyrmion core expansion and relaxation back to the initial state. 
In this regime, the response of the system to the magnetic pulse is not strong enough to excite the meron pair state and switch the polarity of the surrounding magnetization. On the other hand, for pulses that are too long the switched skyrmion appears when the field is still too strong, which causes the skyrmion to collapse, see the white region in Fig.~\ref{fig_switch_diagram_alpha}. For the intermediate pulse width, we observe either successful one-to-one skyrmion switching or the nucleation of a mixed state consisting of iSks and domain walls, similar to the state shown in Fig.~\ref{fig_snapshots}c at $t^\ast=150$~ps. 
The mixed state becomes more apparent for small damping, $\alpha<0.1$, due to the strong influence of the interference between spin waves injected by the free edges.
However, for $\alpha \gtrsim~0.1$, the generation of the spin waves and their interference are significantly suppressed and this results in an enlargement of the successful switching range.
We found that the effect of spin wave interference is also suppressed in the case of a disk shape domain, even for low damping.
In Supplementary Materials \textbf{S3}, we provide the details of the skyrmion switching in a disk-domain together with simulations for varying the values of the exchange coupling constant \textit{J} and show how this affects the field pulse amplitude $B_\textrm{0}$ required for successful switching. 
For realistic coupling parameters we identified a wide switching range in terms of magnetic pulse amplitude $B_\textrm{0}$ and width $t_\textrm{w}$.
We identified the optimal polar angle $\theta$ of the magnetic field pulse for successful switching to be in range between 35$^\circ$ and 55$^\circ$.  
Outside this range, one may observe either the excitation of the breathing mode, the mixed state or skyrmion collapse, while the variation in the azimuthal angle $\varphi$ does not affect significantly the switching mechanism. 
In the present simulations we used $B_\textrm{0}$ = 1 T. However, by adjusting the material parameters, in particular the anisotropy, stable switching can be obtained for lower fields $B_\mathrm{0}\sim 100$ mT but longer pulses $t_\mathrm{w}\sim 100$ ps.

In conclusion, we presented a complete phase diagram for a thin magnetic film, using a discrete model, which includes the DMI and anisotropy energies, that shows stable isolated skyrmion solutions at zero applied field. In particular, we found a collapse line for an isolated skyrmion, which defines its finite parameter-range of existence. The degeneracy of the ground state allows the existence of two skyrmion solutions with mutually inverted spin structure, opposite polarity and topological charge. We demonstrated that the switching between them can be driven by a single inclined magnetic field pulse below 100 ps. The general mechanism of the chiral skyrmions switching follows a sequence of transient topological states: a \textit{chiral-achiral} meron pair, an \textit{achiral} skyrmion and a \textit{half-switched} skyrmion. The newly proposed skyrmion switching mechanism can be achieved in a wide range of material and pulse parameters and allows a repetitive skyrmion toggling, on the GHz scale, which makes these findings of interest for potential applications in MRAM-like devices. 

\bigskip

\textbf{Method}

The total Hamiltonian of a thin film of chiral magnet \cite{Rybakov_15} is given by 
\begin{align}
\mathcal{H}  =   -\sum_{\Braket{i<j}} J \,  (\vcc{n}_i \cdot \vcc{n}_j) - \sum_{\Braket{i<j}}  \vcc{D}_{ij} \cdot  [\vcc{n}_i \times \vcc{n}_j ] 
 - \sum_{i}  K  ({n}^z_i  )^2  - \sum_{i}(\vcc{n}_i \cdot \vcc{b}),       
\label{eq_model-Hamiltonian}
\end{align}
where $\vcc{n}_i=\vcc{m}_i/\mu_\textrm{s}$ is a unit vector of the magnetic moment at lattice site $i$;  
$J$ is the exchange coupling constant; $\vcc{D}_{ij}$ is the Dzyaloshinskii-Moriya vector defined as $\vcc{D}_{ij}=D \vcc{r}_{ij}$, with $D$ a scalar constant and $\vcc{r}_{ij}$ a unit vector pointing from site $i$ to site $j$, see Fig.~\ref{fig_stability}a; $K$ is the out-of-plane uniaxial anisotropy constant, and the last term, $\vcc{b}=\mu_\textrm{s}\vcc{B}$, describes the coupling of the magnetic moments to an external applied field $\vcc{B}$. Here we assume $\mu_\textrm{s}=2\mu_\textrm{B}$.
We restrict ourselves to nearest-neighbor interaction in order to work with a reduced number of variables and conserve generality of the results.

To describe the skyrmion switching dynamics we use atomistic spin dynamic simulations based on the solution of the Landau-Lifschitz-Gilbert equation \cite{Lazaro98}:
\begin{equation}
\label{SLLG}
\tdd{\vcc{n}_i}{t}= -\frac{\gamma}{(1+{\alpha}^2)\mu_\textrm{s}}
 \vcc{n}_i \times ( \vcc{B}_i 
                    +\alpha \vcc{n}_i \times \vcc{B}_i)
\end{equation}
\setlength{\belowdisplayskip}{0pt}  
where $\vcc{B}_i$ is an effective magnetic field defined by $\vcc{B}_i= - \partial{\Ham}/\partial{\vcc{n}_i}$, $\gamma$ is the gyromagnetic ratio and $\alpha$ is a dimensionless damping  coefficient. 
Eq.~(\ref{SLLG}) is solved using currently the most efficient algorithm proposed in Ref.~\onlinecite{SIB} realized in the juSpinX code \cite{JSpinX}.  
We used a time-dependent magnetic field pulse defined by a Gaussian function,
\begin{equation}
\vcc{B}_\textrm{p}(t)  = B_\textrm{0} \exp \left[-\dfrac{ \left(t - t_\textrm{p}\right)^2 }{2 t^2_\textrm{w}} \right]\vcc{\hat{e}}_\textrm{B} 
\label{Gaussian}
\end{equation}
applied into a direction $\vcc{\hat{e}}_\textrm{B}$ inclined relative to the surface normal by a polar angle $\theta$ and an azimuthal angle $\varphi$, as shown in Fig.~\ref{fig_stability}a.
$B_\textrm{0}$, $t_\textrm{w}$ and $t_\textrm{p}$ are the amplitude, Gaussian width and position of the maximum of the pulse, respectively, see inset in Fig.~\ref{fig_stability}a. 
The pulse is assumed to penetrate uniformly through the whole domain. The time step in the simulations is fixed to 1 fs while a typical simulation time is about 1 ns, which is long enough compared to a typical $t_\textrm{w}$ of the order of $10 \textup{--} 30$ ps. 

To describe the topological properties of the solutions and transient spin structures observed within atomistic spin dynamics, we followed Berg and L\"uscher definition for $Q$ on a discrete lattice \cite{Berg81}, for details see Supplementary Materials \textbf{S4}.

\bigskip
\textbf{Acknowledgments}

The authors thank F. N. Rybakov, B. Dupe, M. V. Mostovoy and A. N. Bogdanov for fruitful discussions and acknowledge A. Khajetoorians for critical reading of the manuscript. This work was supported by the European Unions Seventh Framework Program (FP7/2007-2013) FEMTOSPIN, ERC Grant Agreement No. 339813 (EXCHANGE), de Stichting voor Fundamenteel Onderzoek der Materie (FOM) and the Netherlands Organization for Scientific Research (NWO).

\bigskip
\textbf{Contributions}

C.H conceived the project and prepared the initial manuscript draft. 
C.H, N.S.K and A.K.N. carried out the numerical simulations. 
N.S.K and A.K.N carried out the analysis and interpretation of the results, and completed the manuscript revisions. 
T.R and S.B supervised the study. 
All authors discussed the results and contributed to the writing of the paper.

\bigskip

\bigskip
\textbf{Correspondence to:} Nikolai S. Kiselev (n.kiselev@fz-juelich.de).

\newpage


\part*{\Large \centering Supplementary Materials}



  

 

\maketitle

\renewcommand{\thesection}{\text{S}\arabic{section}}
\numberwithin{equation}{section}
\numberwithin{figure}{section}

\setcounter{equation}{0}
\setcounter{figure}{0}

\section{Details of the phase diagram calculation}\label{Supp_stability}

\begin{figure}[h]
 \centering
 \includegraphics[width=16.0cm]{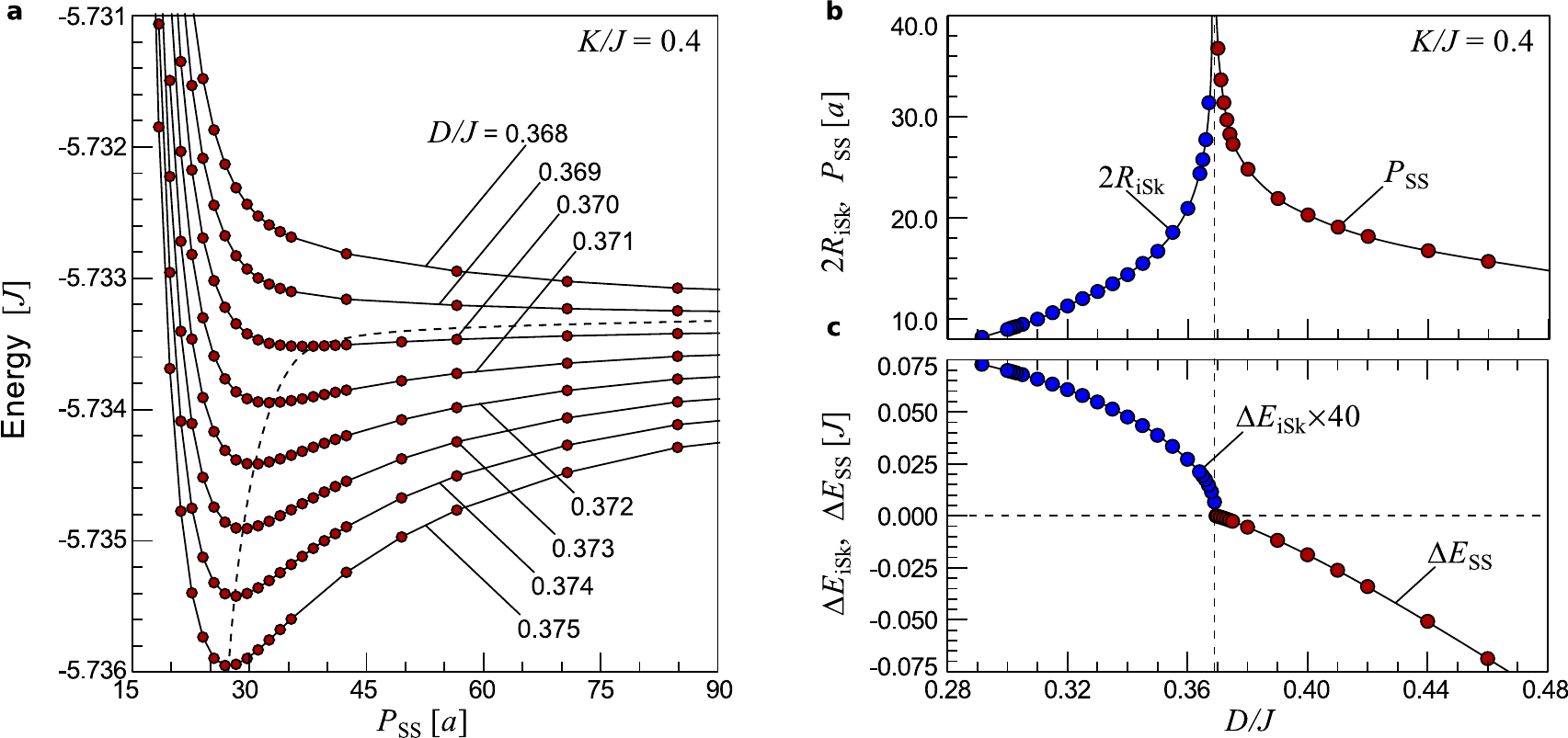}
\caption{\textbf{The scheme to determine the phase transition point for varying $D/J$ at fixed $K/J$:} 
(a) Energy density as a function of spin spiral (SS) period, $P_\textrm{SS}$, numerically calculated at zero applied field for model Hamiltonian Eq.~$1$ in the main text. 
Calculations are done for a fixed ratio $K/J=0.4$, and varying values of $D/J$ close to the phase transition into the saturated ferromagnetic state.
Doted line follows the position of the energy minimum for each $D/J$. 
It approaches (converges to) the energy of the ferromagnetic state when $D/J$ approaches the critical value.
The period of SS is presented in the unit of crystal lattice parameter $a$.
(b) The dependence of the SS period and the size of an isolated skyrmion on $D/J$ for a fixed $K/J = 0.4$. 
Near the phase transition point, $P_\textrm{SS}\rightarrow \infty$ and $D_\textrm{iSk}\rightarrow \infty$.
(c) The energy density for the SS state, $\Delta E_\textrm{SS}$, and for the isolated skyrmion state, $\Delta E_\textrm{iSk}$, relative to the saturated ferromagnetic state as function of $D/J$. 
The equilibrium skyrmion size calculated in square domain of $100 \times 100 \times 3$ spins with periodical boundary conditions in the $xy$-plane.}
\label{fig_ssperiod}
\end{figure}

In the main text, we presented the phase diagram of the ground state of an extended three monolayer thick film, see Fig.~1a.
Here, we present the details of the calculations leading to the phase diagram as well as the dependence of the phase transition lines on the film thickness.
To find the phase transition line between spin spiral (SS) and saturated ferromagnetic (FM) states we used the following scheme. 
We numerically calculated an equilibrium period and a corresponding energy for varying values of $D/J$ at fixed $K/J$. 
The value at which the period of the SS tends to infinity and the energy difference between SS and FM states tends to zero, we identify as second order phase transition line.
In Fig.~\ref{fig_ssperiod} we show an example of such a calculation for $K/J=0.4$.
In particular, Fig.~\ref{fig_ssperiod}a shows the total energy density as a function of the SS period $P_\textrm{SS}$ for different values of $D/J$.
It is easy to show that in case of a thin film of a chiral magnet with a simple cubic structure and nearest-neighbor exchange and Dzyaloshinskii-Moriya interactions (DMI) the preferable direction for the SS propagation is $\braket{110}$, along the diagonal of a plane of the elementary cubic cell.
The total energy density is calculated in unit of exchange coupling constant $J$. 
The position of the energy minimum for varying $D/J$ denoted by the doted line in Fig.~\ref{fig_ssperiod}a corresponds to the equilibrium period of the SS state.
As follows from Fig.~\ref{fig_ssperiod}a and b the equilibrium period runs to infinity between $D/J$=0.369 and 0.370 while the energy of SS tends to the energy of the FM state, which can be calculated precisely and for this case equals $-$5.73(3) $J$.   
In Fig.~\ref{fig_ssperiod}c we show the energy density difference between SS and FM states. The $\Delta E_\textrm{SS}$ tends to zero when one approaches the phase transition point at about $D/J$ = 0.3695 $\pm$ 0.0005.
Such an approach allows to identify the phase transition point for $D/J$ with a precision, which depends on the step size in $D/J$.

On the left side of the transition line in Fig.~\ref{fig_ssperiod}b, we also show the size dependence for the metastable isolated skyrmion (iSk).
In Fig.~\ref{fig_ssperiod}c, the calculated energy of an iSk with respect to the FM state illustrates the metastability of iSk, since the energy of the iSk is always higher than the energy of FM state.

\begin{figure}[!t]
 \centering
 \includegraphics[width=15.0cm]{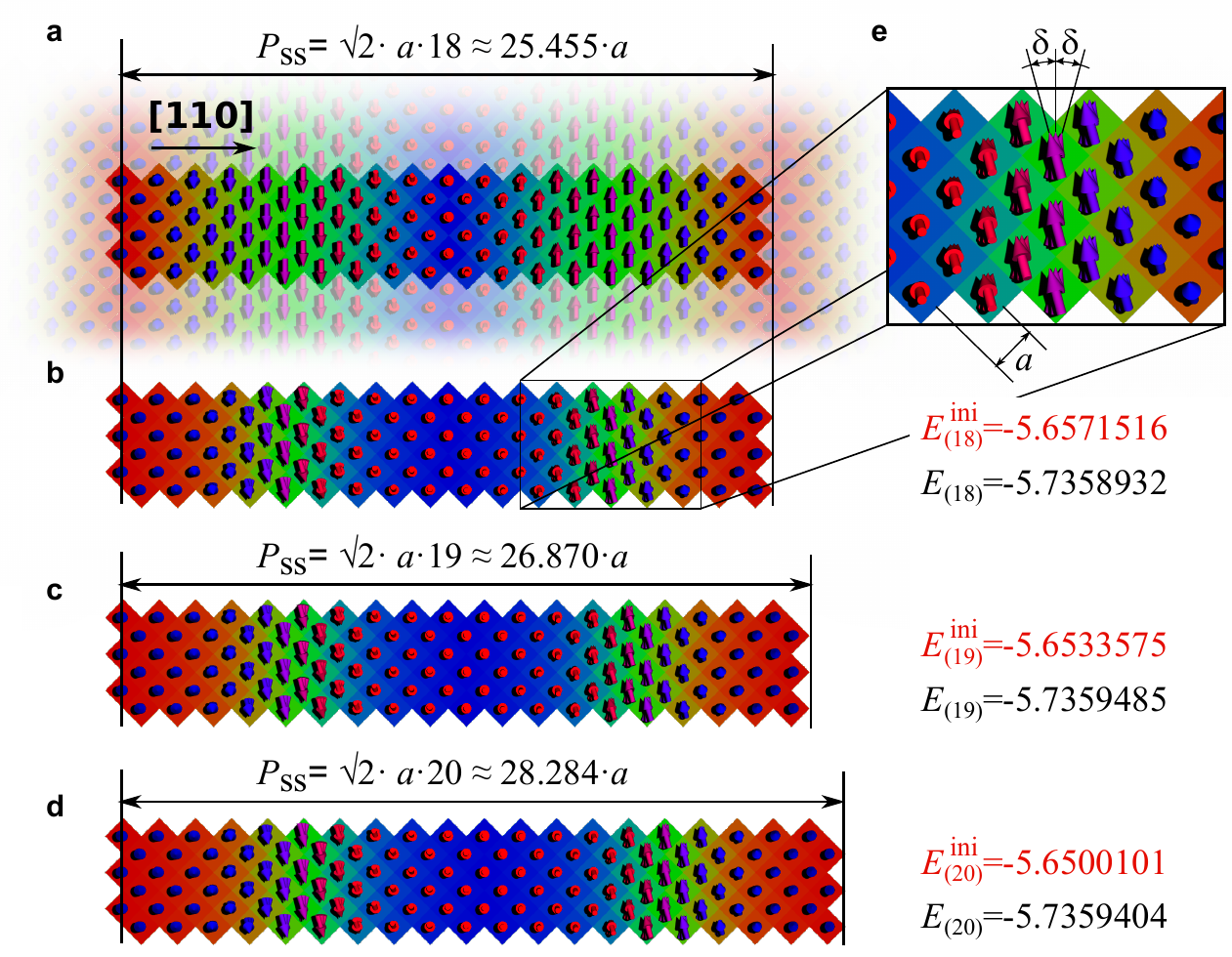}
\caption{\textbf{Snapshots of SS propagating along $\braket{110}$ of a simple cubic lattice for different characteristic domain size defined as $\sqrt{2} \cdot a \cdot N$ along $\braket{110}$:}  
(a) Homogeneous SS used as the initial state for the following relaxation. 
Transparent spins indicate periodical boundary conditions used in the relaxation process.
(b) Inhomogeneous SS after the relaxation of the initial state shown in a.  
(c)-(d) Inhomogeneous SS states after the relaxation for the characteristic domain size with $N=19$ and $20$, respectively.  
(e) Zoomed view of the inhomogeneous SS in b. The angles $\delta$ indicate a modulation of the spins along the thickness.
$E^\textrm{ini}_{(N)}$ (in red) indicates the energy density in the unit of $J$ for the homogeneous initial SS state and $E_{(N)}$ (in black) indicates the energy density of the inhomogeneous fully relaxed SS state.   
}
\label{fig_ssperiod_images}
\end{figure}

To find the dependence of the energy density on the period of the SS presented in Fig.~\ref{fig_ssperiod}a, we performed the following calculations.
We calculated the energy of the SS state after a full relaxation on a domain with periodic boundary conditions (PBC).
As an initial configuration we used a homogeneous SS with the period defined as $P_\textrm{SS}=N/i$, where $N$ is the size of the domain along $\braket{110}$ and $i$ is a fixed integer number.
Due to PBC, the period of the SS during the relaxation remains conserved.
By varying the size of the simulated domain, one can find the energy dependence as a function of $P_\textrm{SS}$. 
Fig.~\ref{fig_ssperiod_images} illustrates our approach for the case $K/J=0.4$ and $D/J=0.375$.
Here we assume $i=1$, the size of the domain corresponds to one SS period.
Fig.~\ref{fig_ssperiod_images}a and b show the initial homogeneous and relaxed inhomogeneous SS states, respectively.
Figs.~\ref{fig_ssperiod_images}c and d show relaxed SS for larger domains corresponding to larger SS periods.
Compare the relaxed energies corresponding to Figs.~\ref{fig_ssperiod}b, c and d in this particular case, the lowest energy state is the SS state with period $P_\textrm{SS}=\sqrt{2}a\cdot19=26.87a$.
The step in energy dependence on the period in such an approach is inversely proportional to the integer number of periods $i$ in the domain, $\Delta P_\textrm{SS}=\sqrt{2}a/i$.
However, for higher precision it is important to use large $i\gtrsim 10$, when the equilibrium period of the spin spiral is incommensurate with the lattice constant $a$.

Note, there is an analytical solution for the phase transition line between ferromagnetic and spin spiral states derived by Dzyaloshinskii in the frame of  micromagnetic continuum approximation \cite{Dzyaloshinskii, Izumov84}.
However the approach of Dzyaloshinskii ignores the effect of magnetization modulation along the thickness.
Thus, it is valid only for two limiting cases: pure two-dimensional case (single monolayer) and the bulk system where the effects of the chiral surface twist can be neglected.  
Recently, this effect and the role of magnetic modulation along the thickness of the magnetic layer has been studied in the context of skyrmion stability in cubic helimagnets \cite{Rybakov_15}.
It has been shown that such modulations are localized close to the free surfaces and sufficiently reduce the energy of the skyrmion state as well as lead to the stabilization of other earlier unknown localized magnetic states {--} \textit{chiral bobbers}.

Iin both cases, the iSk and the spin spiral, we observe pronounced effects of modulations along the thickness in both cases of iSk and spin spiral.
Fig.~\ref{fig_ssperiod_images}e shows the zoomed in area of the relaxed spin spiral state and illustrates the effect of chiral twist along the thickness of the layer, see small angle $\delta$ between spins in the middle layer and spins in top and bottom layers.
A similar modulation is apparent in the spin structure of iSk.  

\begin{figure*}[!t]
 \centering
 \includegraphics[width=17.0cm]{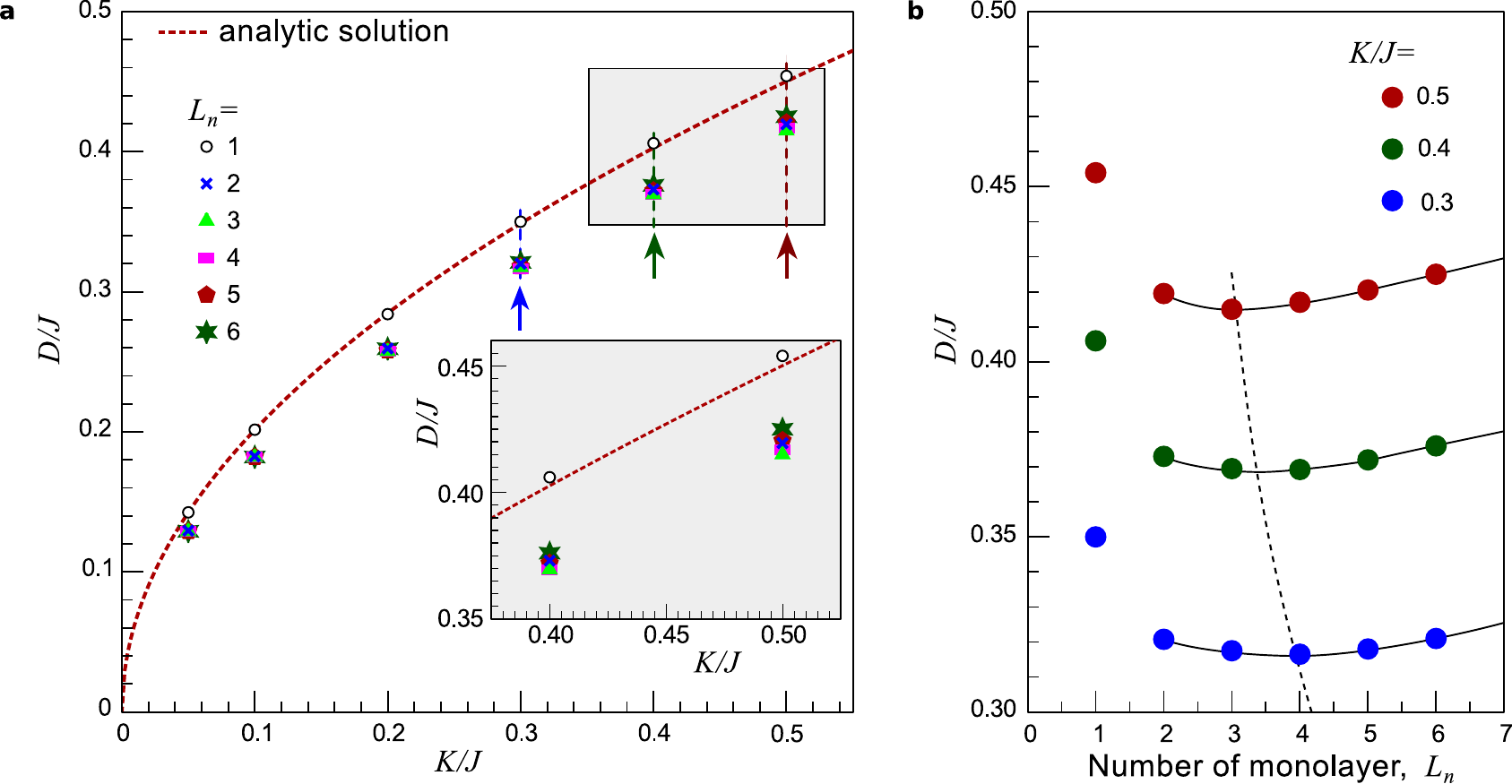}
\caption{\textbf{Phase transition of spin spiral state for different thicknesses:} 
(a) Phase transition between spin spiral and ferromagnetic states calculated for different thicknesses of the $(001)$ film of a simple cubic lattice.
Dashed line is the analytical solution (Eq.~\ref{criterion}) employing the micromagnetic approach.
Here $L_{n}$ denotes the thickness of the film in number of monolayers.
Inset shows zoomed area marked as (gray) rectangle in the main figure.
(b) Dependence of the critical values of $D/J$ for the phase transition as a function of the magnetic film thickness $L_{n}$ for fixed values of $K/J$ marked by arrows in figure a.}
\label{fig_transition}
\end{figure*}

Such modulations significantly reduce the energy of inhomogeneous states.
In order to illustrate the pronounced effects of modulations on the stability of a spin spiral, we adopt an analytical solution derived by Dzyaloshinskii and compare the results with our numerical calculations.
In particular, in the micromagnetic approach assuming small angles of rotation between nearest neighbor spins, the energy density of inhomogeneity for the spin spiral with period $\lambda$ is
\begin{equation}
\Delta E=
\dfrac{1}{\lambda}\int_{0}^{\lambda} 
\left[ 
\mathcal{K}\cdot\sin^2\phi + \mathcal{J}\dfrac{\textrm{d}\phi}{\textrm{d}r}^2 - \mathcal{D}\dfrac{\textrm{d}\phi}{\textrm{d}r}
\right]\textrm{d}r,
\label{energy_ss}
\end{equation} 
where $\phi \equiv \phi(r)$ describes the polar angle of magnetization. 
For a simple cubic lattice, the micromagnetic constants in Eq.~\ref{energy_ss} are related to the constants of a discrete model via
$\mathcal{J}=J/a $, and $\mathcal{D}=2D/a^2$,
$\mathcal{K}=K/a^3$. 
Minimization of Eq.~\ref{energy_ss} gives an expression for energy minimum $\Delta E_\textrm{min}$ and an equilibrium period of the spin spiral $\lambda_\textrm{min}$.
Ref.~\cite{Bogdanov_89} gives a simple and elegant proof for the following inequality
\begin{equation}
\Delta E_\textrm{min} \geq 8 \sqrt{\mathcal{K}\mathcal{J}}\pm 2\pi\mathcal{D},
\label{E_min}
\end{equation}
where the first term represents twice the energy of an ordinary Bloch wall, while the second one reflects the contribution of DMI.
The sign of the last term in Eq.~\ref{E_min} depends on the sense of rotation of the spins in the spiral. 
If the value of $\Delta E_\textrm{min} > 0$, the ground state corresponds to a collinear FM state, if $\Delta E_\textrm{min} < 0$ one may conclude that there exists an inhomogeneous state, which has an energy lower than the FM state. 
Therefore, a criterion for appearance of an inhomogeneous spin spiral reads
\begin{equation}
\frac{\mathcal{D}}{\mathcal{J}}=
\frac{4}{\pi}\sqrt{\frac{\mathcal{K}}{\mathcal{J}}},
\end{equation}
or in units of the discrete model for the case of a simple cubic lattice
\begin{equation}
\frac{D}{J}=
\frac{2}{\pi}\sqrt{\frac{K}{J}}.
\label{criterion}
\end{equation}

In Fig.~\ref{fig_transition}a we present the analytical solution for the phase transition line (Eq.~\ref{criterion}), see the dashed red line.
The symbols in the figure represent our numerical calculations for different thicknesses of the film. 
As has been mentioned before, the micromagnetic approach is a good approximation for transitions in the case of one monolayer, see $L_n=1$, open circles. 
The discrepancy appears only for the high anisotropy case, $K/J>0.3$, see inset in Fig.~\ref{fig_transition}a.
This is because in the high anisotropy case, the approximation of small angle rotation fails. 
For any finite thickness $L_n > 1$, the transition always occurs below the transition corresponding to the single monolayer ($L_n=1$).
Indeed, Fig.~\ref{fig_transition}b shows an abrupt change in the transition values of $D/J$ between pure two-dimensional case, $L_n=1$ and the one of finite thicknesses, $L_n \geq 2$.
The transition value of $D/J$ passes through a minima and then gradually increases.
The critical thickness which corresponds to the minima strongly depends on anisotropy value, see the dashed line in Fig.~\ref{fig_transition}b.
For the bulk limit, the critical value of $D/J$ gradually converges to the pure two-dimensional case of single monolayer.
Such a dependence for the transition line on the thickness has to be taken into account for the calculation of the stability range for iSks. 

\section{Size of isolated skyrmion on a discrete lattice} \label{subsection_size-iSk}

\begin{figure}[hbt]
 \centering
 \includegraphics[width=8cm]{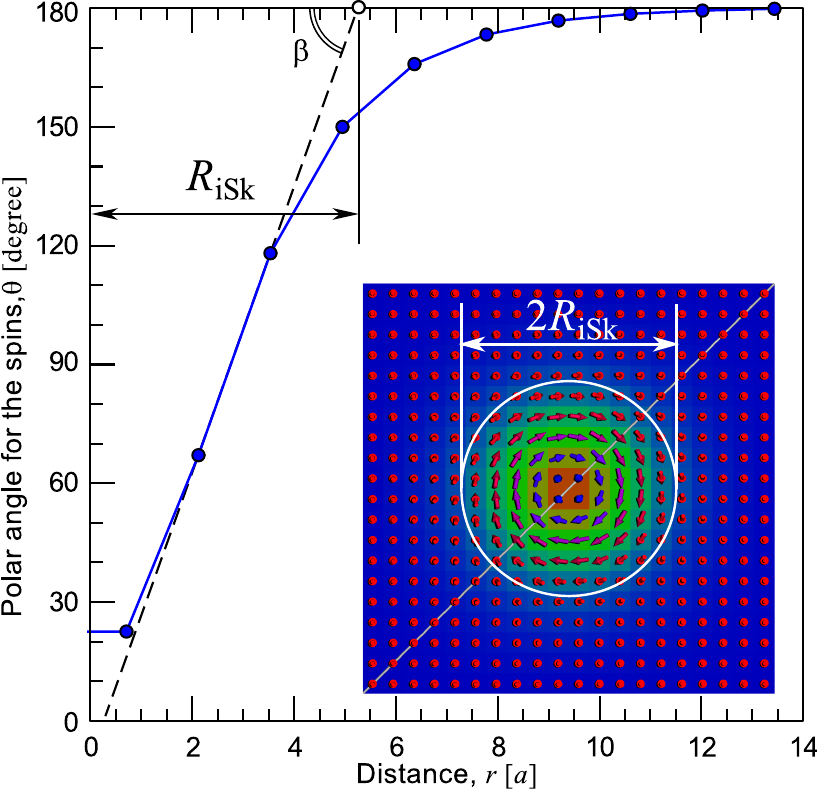}
 \caption{\textbf{Size of an isolated skyrmion:} Polar angle dependence, $\theta(r)$, for spins along the profile of magnetic skyrmion shown in the inset. $r=0$ denotes the center of the skyrmion. Doted line shows tangent to the discrete skyrmion profile. Color of the arrows denote the direction of z-component of the magnetization (blue corresponds to up, red to down).}
 \label{fig_skyrmion_size}
\end{figure}

Figure~\ref{fig_skyrmion_size} illustrate our approach defining the size of an iSk for a discrete model. 
The tangent line for the skyrmion profile fitted by a linear function at the point of inflection, \textit{i.e.} with largest inclination angle $\beta$, see the straight doted line running through two nearest points.
Diameter of the skyrmion is defined as twice the distance between the center of a skyrmion and the intersection point of the tangent with the $x$-axis, \textit{i.e.} R$_{iSk}$. 

\section{Skyrmion switching dynamics in disk shape domain} \label{subsection_disk_domain}

\begin{figure}[!t]
 \centering
  \includegraphics[width=17.0cm]{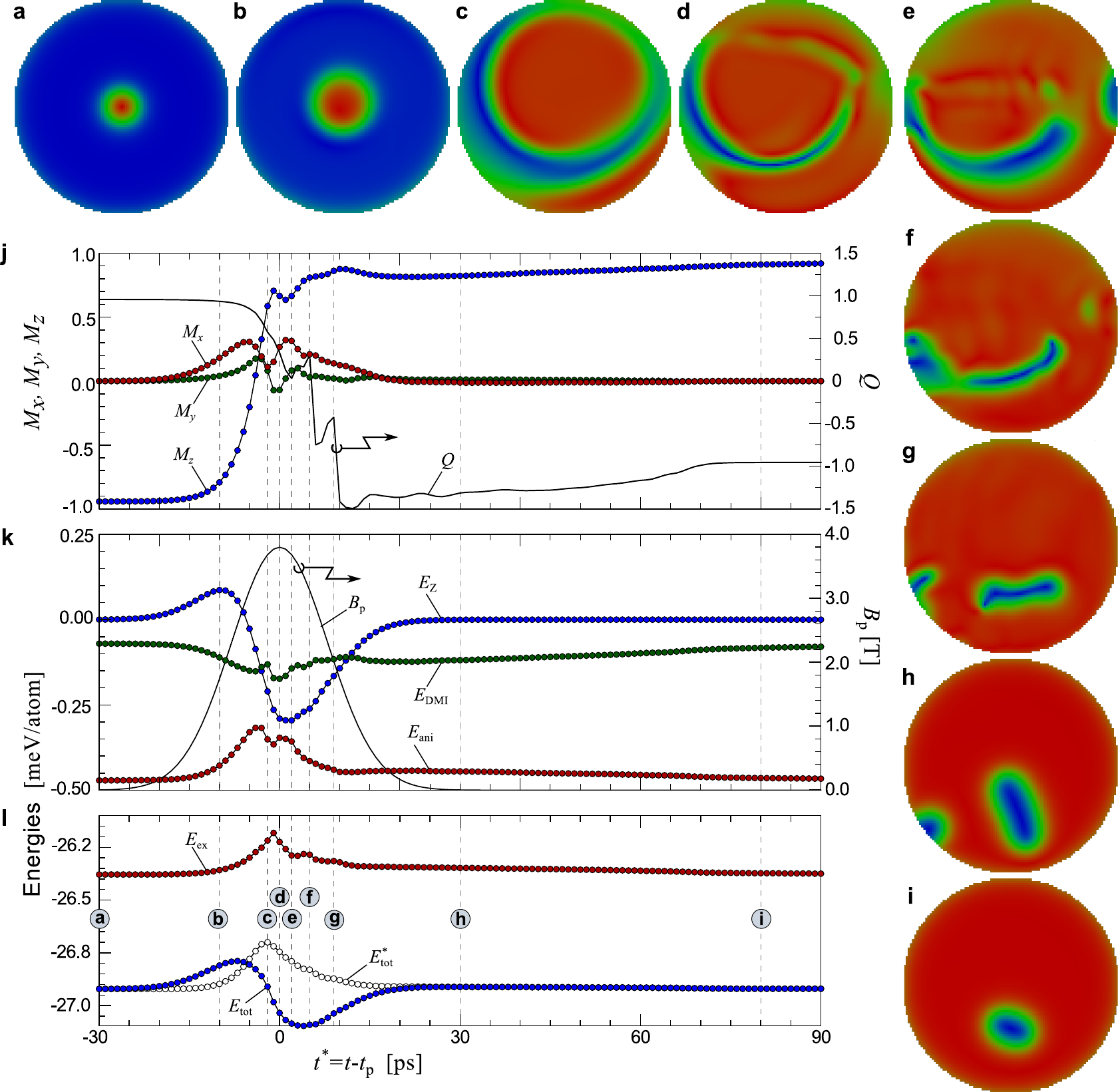}
 \caption{\textbf{Dynamics of skyrmion switching by a short inclined magnetic field pulse in disk domain geometry:} (a)-(i) Snapshots of the spin structure taken at time denoted by vertical dotted lines in the panels (j)-(l) with corresponding labels.
Here, $M_\textrm{x}$, $M_\textrm{y}$, $M_\textrm{z}$ are the components of average  magnetization and the topological charge \textit{Q} is calculated using Eq.~\ref{eq_top_charge},
$E_\textrm{ex}$, $E_\textrm{DMI}$, $E_\textrm{ani}$, and $E_\textrm{Z}$ are energy contributions of exchange, DMI, anisotropy, and Zeeman energy, respectively.
$B_\textrm{p}$ exhibits the profile of magnetic field pulse.
Note, $E_\textrm{tot}$ is the sum of all energy terms while $E^*_\textrm{tot}=E_\textrm{tot}-E_\textrm{Z}$.
}
 \label{fig_switch_disk}
\end{figure}

\begin{figure}[hbtp]
 \centering
  \includegraphics[width=17.0cm]{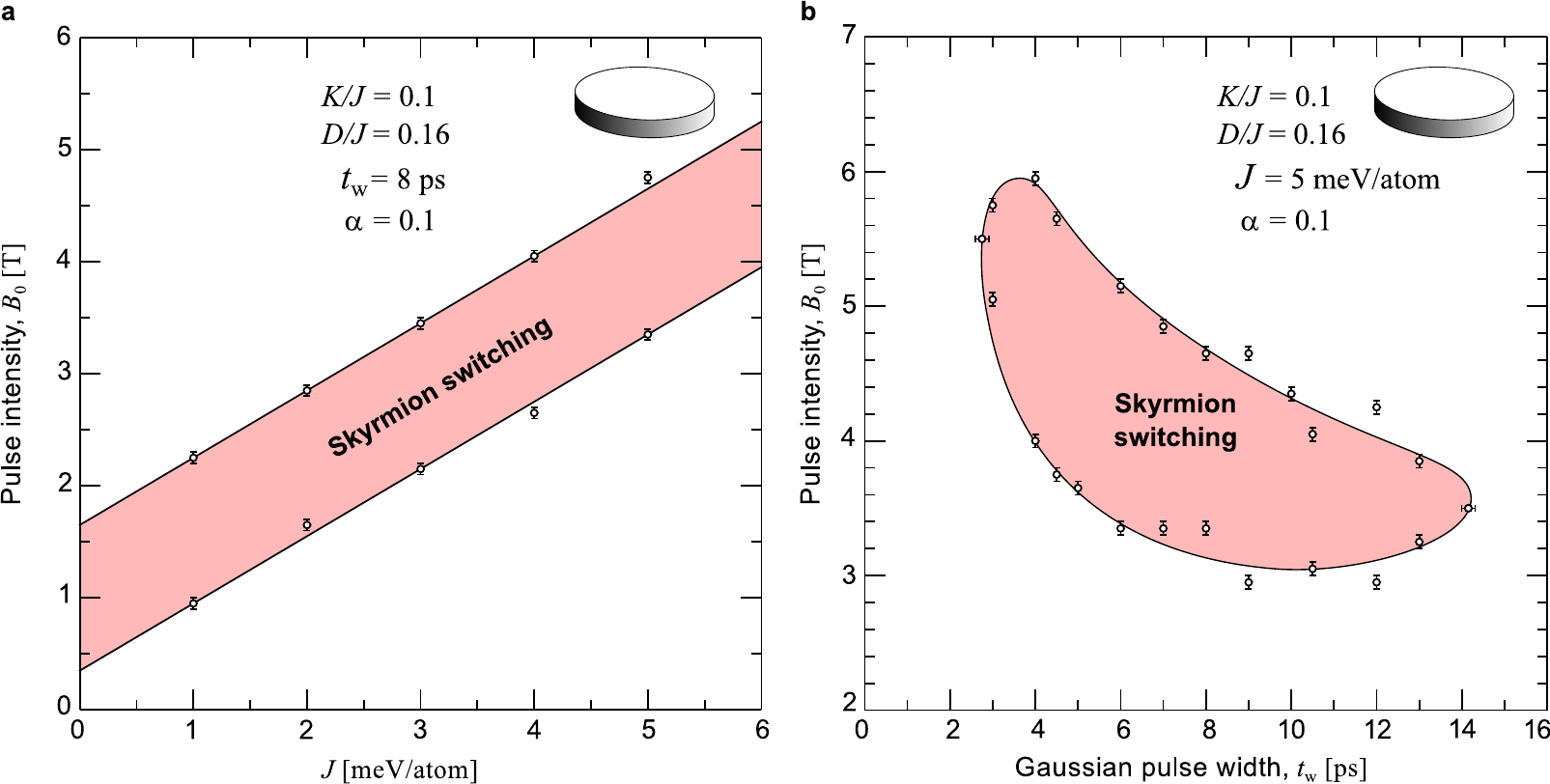}
 \caption{\textbf{Switching phase diagrams using disk domain:}
(a) Switching diagram calculated in terms of intensity of magnetic field pulse $B_0$ and exchange stiffness $J$. 
(b) Switching diagram calculated for varying intensity of Gaussian magnetic field pulse $B_0$ and pulse duration $t_\textrm{w}$.
The values for the fixed parameters (coupling constants, applied pulse angle, and damping)  are displayed in the figures.
Both switching diagrams are calculated for disk domain with diameter and thickness of 100 spins and 3 monolayers, respectively.
}
 \label{fig_switch_diagram_disk}
\end{figure}

Our investigations on the topological dynamics and energetics of skyrmion switching have been continued for a disk shape domain with open boundary condition (OBC).
The considered diameter and thickness of the disk are 100 and 2 atomic distances (three monolayer), respectively.
We have used a realistic set of parameters including absolute values of exchange $J$ = 5 meV, relative values of DMI $D/J$ = 0.16 and anisotropy $K/J$ = 0.1, $\mu_\textrm{s}$ = 2$\mu_\textrm{B}$ and damping $\alpha$ = 0.1.
An inclined magnetic Gaussian pulse with polar angle $\theta$ = 45$^\circ$ penetrates uniformly through the disk. 
The pulse has an intensity $B_0$ = 4 T and a Gaussian width $t_\textrm{w}$ = 8 ps.
In Fig.~\ref{fig_switch_disk}, the snapshots a - i display the sequence of states taken at different moments in time (see the vertical dashed lines in figures j - l) through which the system passes during the switching.
Note that the topological charge of skyrmions in OBC is an ill-defined quantity.
Thereby, \textit{Q} for the initial state is found to be +0.93 and $-$0.93 after the switching, see the time dependence of \textit{Q} presented in Fig.~\ref{fig_switch_disk}j.
Similar to the switching mechanism discussed in the main text, the simultaneous effects of skyrmion expansion and change in the polarity of the surrounding ferromagnetic state result in the formation of an vortex-antivortex pair with distorted shapes, see snapshot c.
Injected spin waves from the free boundary start to propagate and interact with the excited skyrmion core.
Comparing the switching processes in the square and disk domains, one may conclude that the effect of spin-wave interference in the disk shape domain is less pronounced, compare the snapshots in Fig.~2d in the main text and the snapshots in Fig.~\ref{fig_switch_disk}.
After some simulation time at about $t^\ast =$ 5 ps, a localized achiral skyrmion appears, while the magnetization of the surrounding state is almost flipped, see snapshot f.
A sharp fall in \textit{Q}, which appears between the time step marked with f and g represents the transition from an achiral skyrmion to a half-switched skyrmion.
Contrary to the PBC, here for the case of OBC, the half-switched skyrmion may appear with nonzero topological charge, see the time variation of \textit{Q} in Fig.~\ref{fig_switch_disk}j at about $t^\ast =$ 9 ps.
The nonzero value of \textit{Q} is due to the presence of domain attached to the edge, see the snapshot g.
The transition to a chiral skyrmion, see snapshot h, is further accompanied by another sharp jump of \textit{Q} between $t^\ast = $ 9 and 10 ps.
Due to the presence of the extra domain, the absolute value of \textit{Q} is larger than unity.
Such domain corresponds to an unstable state.
After the relaxation it disappears and the topological charge converges close to unity, see the time dependence of \textit{Q} between $t^\ast$ = 30 and 80 ps and the corresponding snapshots.

The average magnetization and the energy contributions presented in Fig.~\ref{fig_switch_disk} show qualitatively the same time dependencies as those in the case of square domain, see Fig.~$3$ in the main text. 
Additional distortions in time dependencies are expected due to the OBC.

It is important to note that the skyrmion switching turned to be robust within a wide range of parameters.
We have investigated the skyrmion switching by varying the absolute values of exchange coupling, pulse width and intensity.
The range for successful switching is presented in the switching diagrams, Fig.~\ref{fig_switch_diagram_disk}.
The major energy contribution in the switching energetics comes from the exchange interaction $J$, which is usually about one order of magnitude higher than all other interactions such as $D$ and $K$.
In Fig.~\ref{fig_switch_diagram_disk}a, a parameter domain for successful switching is identified varying pulse intensity $B_0$ and the absolute values of $J$ for fixed relative values of $K/J$ and $D/J$.
Here, the damping parameter $\alpha$ = 0.1, Gaussian width $t_\textrm{w}$ = 8 ps and polar angle $\theta$ = 45$^\circ$ have been used.
The successful switching regime (green area) in terms of $B_0$ is sufficiently wide, of about 1.2 T, while the lower and upper critical fields increase linearly with exchange coupling.

Thereafter, using a realistic value of exchange interaction $J$ = 5 meV, we calculated the switching diagram in the parameter space of $B_0$ and $t_\textrm{w}$ presented in Fig.~\ref{fig_switch_diagram_disk}b.
A wide area corresponding to \textit{one-to-one} successful switching is identified existing within the range about 3 to 6 T and $t_\textrm{w}$s between 3 to 14 ps.
The rough edge in the switching area reflects the complex energy landscape of the system with a large number of metastable states while inside this region we always find stable one-to-one skyrmion switching.

\section{Definition of topological charge on a discrete lattice} \label{Supp_top-charge}

\begin{figure}[!t]
 \centering
 \includegraphics[width=8.0cm]{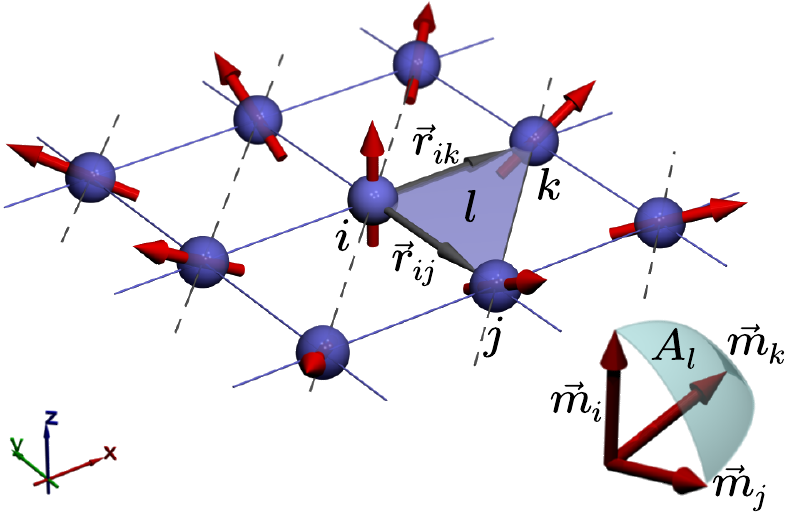}
 \caption{\textbf{Topological charge on a discrete lattice:} Two-dimensional square lattice partitioned into elementary triangles. Solid angle $A_l$ based on three magnetic moments $\vcc{m}_i$, $\vcc{m}_j$, $\vcc{m}_k$ located at the vertices of an elementary triangle $l$, marked by blue shading.}
 \label{fig_top_charge}
\end{figure}

The discrete model requires a proper definition of the topological charge on a lattice of spins $\vcc{m}(x_i,y_i)$, where $i$ runs over all the lattice sites.
We follow the definition given by Berg and L\"uscher \cite{Berg81} and arrive at the following expression:
\begin{equation}
Q= \frac{1}{4\pi}\sum_l A_l,
\label{eq_top_charge}
\end{equation}
with
\begin{equation}
\cos\left(\frac{A_l}{2}\right)=\frac{1  + \vcc{m}_i \cdot \vcc{m}_j + \vcc{m}_j \cdot \vcc{m}_k + \vcc{m}_k \cdot \vcc{m}_i}
{\sqrt{2\left(1+\vcc{m}_i \cdot \vcc{m}_j\right)\left(1+\vcc{m}_j \cdot \vcc{m}_k\right)\left(1+\vcc{m}_k \cdot \vcc{m}_i\right)}}
\label{A_l}
\end{equation}
where $l$ runs over all elementary triangles defined on the square lattice; $A_l$ is the area of the spherical triangle with vertices $\vcc{m}_i$, $\vcc{m}_j$, $\vcc{m}_k$, see Fig.~\ref{fig_top_charge}.
The sign of $A_l$  is determined as sign$\left(A_l\right)=$ sign $\left[\vcc{m}_i\cdot\left(\vcc{m}_j\times\vcc{m}_k\right)\right]$.

The vertices $i$, $j$, $k$  of each elementary triangle are numbered in a counter-clockwise sense relative to the surface normal vector $\vcc{n}$ pointing in positive direction of the $z$-axis.
The latter means that the numbering should satisfies the following condition $\vcc{n}\cdot(\vcc{r}_{ij}\times \vcc{r}_{ik})>0$, where $\vcc{r}_{ij}$ is a connection vector directed from lattice site $i$ to $j$.

The parameter $a_l=A_l/4\pi$ can be thought as \textit{local topological charge}, which takes values in the range of $-0.5 < a_l < +0.5$.
Note, according to Berg and L\"uscher, \cite{Berg81} there is a set of \textit{exceptional} spin configurations for which \textit{Q} is not defined but still \textit{measurable} as $A_l$ in Eq.~(\ref{A_l}) is defined for all possible spin configurations.

\section{Supplementary movie} \label{subsection_switching-video}
The movie shows the back and forth skyrmion switching by the sequence of magnetic field pulses of amplitude $B_\textrm{0}$ = 3 T and width $t_\textrm{w}$ = 15 ps with alternating polar angles $\theta=45^\circ$ and $\theta^\prime=135^\circ$. 
For the simulations in a square shape domain of $100 \times 100 \times 3$ spins, we have used the material parameters $J=5$~meV, $D/J=0.16$, $K/J=0.1$, $\alpha=0.1$ and OBC.
An interval of about 200 ps between two consecutive switching events suggests a skyrmion switching rate in the GHz range.
See \href{https://youtu.be/KQ1DSBsTOak}{\textbf{www.youtube.com}} ({\footnotesize https://youtu.be/KQ1DSBsTOak})


\newpage

\begin{thebibliography}{99}

\bibitem{Dzyaloshinskii}
Dzyaloshinskii, I. E. Theory of helicoidal structures in antiferromagnets. III, \textit{Sov. Phys. JETP} \textbf{20}, 665-668 (1965).

\bibitem{Moriya}
Moriya, T. Anisotropic superexchange interaction and weak ferromagnetism. \textit{Phys. Rev.} \textbf{120}, 91-98 (1960).

\bibitem{TopHall}
Franz, C., Freimuth, F., Bauer, A., Ritz, R., Schnarr, C., Duvinage, C., Adams, T., Bl\"{u}gel, S., Rosch, A., Mokrousov, Y. \& Pfleiderer, C. Real-space and reciprocal-space Berry phases in the Hall effect of Mn$_{1-x}$Fe$_x$Si. \textit{Phys. Rev. Lett.} \textbf{112}, 186601 (2014).

\bibitem{SkHall}
Zang, J., Mostovoy, M., Han, J. H. \& Nagaosa, N. Dynamics of skyrmion crystals in metallic thin films. \textit{Phys. Rev. Lett.} \textbf{107}, 136804 (2011).

\bibitem{Schulz_12}
Schulz, T., Ritz, R., Bauer, A., Halder, M., Wagner, M., Franz, C., Pfleiderer, C., Everschor, K., Garst, M. \& Rosch, A. Emergent electrodynamics of skyrmions in a chiral magnet. \textit{Nature Phys.} \textbf{8}, 301-304 (2012).

\bibitem{Jonietz_10}
Jonietz, F., M\"{u}hlbauer, S., Pfleiderer, C., Neubauer, A., M\"{u}nzer, W., Bauer, A., Adams, T., Georgii, R., B\"{o}ni, R., Duine, R. A., Everschor, K., Garst, M. \& Rosch, A. Spin transfer torques in MnSi at ultralow current densities. \textit{Science} \textbf{ 330}, 1648-1651 (2010).

\bibitem{Bogdanov_89}
Bogdanov, A. N. \& Yablonskii, D. A. Thermodynamically stable "vortices" in magnetically ordered crystals. The mixed state of magnets. \textit{Sov. Phys. JETP} \textbf{68}, 101-103 (1989).

\bibitem{Bogdanov_94}
Bogdanov, A. \& Hubert, A. Thermodynamically stable magnetic vortex states in magnetic crystals. \textit{J. Magn. Magn. Mater.} \textbf{138}, 255-269 (1994).

\bibitem{Bogdanov_99}
Bogdanov, A. \& Hubert, A. The stability of vortex-like structures in uniaxial ferromagnets. \textit{J. Magn. Magn. Mater.} \textbf{195}, 182-192 (1999).

\bibitem{Bogdanov_06}
R\"{o}{\ss}ler, U. K., Bogdanov, A. N. \& Pleiderer, C. Spontaneous skyrmion ground states in magnetic metals. \textit{Nature} \textbf{442}, 797-801 (2006).

\bibitem{Bogdanov_11}
R\"{o}{\ss}ler, U. K., Leonov,  A. A. \& Bogdanov, A. N. Chiral Skyrmionic matter in non-centrosymmetric magnets. \textit{J. Phys. Conf. Ser.} \textbf{303}, 012105 (2011).

\bibitem{Bogdanov_14}
Wilson, M. N., Butenko, A. B., Bogdanov, A. N. \& Monchesky, T. L. Chiral skyrmions in cubic helimagnet films: The role of uniaxial anisotropy. \textit{Phys. Rev. B} \textbf{89}, 094411 (2014).

\bibitem{Muhlbauer_09}
M\"{u}hlbauer, S., Binz, B., Jonietz, F., Pfleiderer, C., Rosch, A., Neubauer, A., Georgii, R. \& Boni, P. Skyrmion lattice in a chiral magnet. \textit{Science} \textbf{323}, 915-919 (2009).

\bibitem{Yu_10}
Yu, X. Z., Onose, Y., Kanazawa, N., Park, J.H., Han, J. H., Matsui, Y., Nagaosa, N. \& Tokura, Y. Real space observation of a two-dimensional skyrmion crystal. \textit{Nature} \textbf{465}, 901-904 (2010).

\bibitem{Yu_11}
Yu, X. Z., Kanazawa, N., Onose, Y., Kimoto, K., Zhang, W. Z., Ishiwata, S., Matsui, Y. \& Tokura, Y. Near room-temperature formation of a skyrmion crystal in thin-films of the helimagnet FeGe. \textit{Nature Mater.} \textbf{10}, 106-109 (2011).

\bibitem{Romming_13}
Romming, N., Menzel, M., Hanneken, C., Bickel, J. E., Wolter, B., von Bergmann, K., Kubetzka, A. \& Wiesendanger, R. Writing and deleting single magnetic skyrmions. \textit{Science} \textbf{341}, 636-639 (2013).

\bibitem{Yu_12}
Yu, X.Z., Kanazawa, N., Zhang, W.Z., Nagai, T., Hara, T., Kimoto, K., Matsui, Y., Onose, Y. \& Tokura, Y. Skyrmion flow near room temperature in an ultralow current density. \textit{Nature Commun.} \textbf{3}, 988 (2012).

\bibitem{Kiselev_11}
Kiselev, N. S., Bogdanov, A. N., Sch\"{a}fer, R. \& R\"{o}{\ss}ler, U. K. Chiral skyrmions in thin magnetic films: new objects for magnetic storage technologies? \textit{J. Phys. D: Appl. Phys.} \textbf{44} 392001 (2011).

\bibitem{Fert_13}
Fert, A., Cros, V. \& Sampaio, J. Skyrmions on the track. \textit{Nature Nanotech.} \textbf{8}, 152-156 (2013).

\bibitem{Parkin_15}
Parkin, S. \& Yang, S-H. Memory on the racetrack. \textit{Nature Nanotech.} \textbf{10}, 195-198 (2015).

\bibitem{MRAM}
Parkin, S. S. P., Roche, K. P., Samant, M. G., Rice, P. M., Beyers, R. B., Scheuerlein, R. E., O'Sullivan, E. J., Brown, S. L., Bucchigano, J., Abraham, D. W., Lu, Y., Rooks, M., Trouilloud, P. L., Wanner, R. A. \& Gallagher, W. J. Exchange-biased magnetic tunnel junctions and application to nonvolatile magnetic random access memory (invited). \textit{J. Appl. Phys.} \textbf{85}, 5828-5833 (1999).

\bibitem{Rybakov_13}
Rybakov, F. N., Borisov, A. B. \& Bogdanov, A. N. Three-dimensional skyrmion states in thin films of cubic helimagnets. \textit{Phys. Rev. B} \textbf{87}, 094424 (2013).

\bibitem{Rybakov_15}
Rybakov, F. N., Borisov, A. B., Bl\"{u}gel, S., \& Kiselev, N. S.  New type of stable particlelike states in chiral magnets. \textit{Phys. Rev. Lett} \textbf{115}, 117201 (2015).

\bibitem{Waeyenberge_06}
Van Waeyenberge, B., Puzic, A., Stoll, H., Chou, K.W., Tyliszczak, T., Hertel, R, F\"{a}hnle, M., Br\"{u}ckl, H., Rott, K., Reiss, G., Neudecker, I., Weiss, D., Back, C.H. \& Sch\"{u}tz, G. Magnetic vortex core reversal by excitation with short bursts of an alternating field. \textit{Nature} \textbf{444}, 461-464 (2006).

\bibitem{Lazaro98}
Garc\'ia-Palacios J. L. \& L\'{a}zaro, F. J. Langevin-dynamics study of the dynamical properties of small magnetic particles. \textit{Phys. Rev. B} \textbf{58}, 14937-14958 (1998).

\bibitem{SIB}
Mentink, J. H., Tretyakov, M. V., Fasolino, A., Katsnelson, M. I. \& Rasing, Th. Stable and fast semi-implicit integration of the stochastic Landau-Lifshitz equation. \textit{J. Phys.: Condens. Matter} \textbf{22}, 176001 (2010).

\bibitem{JSpinX}
Bauer, D. S. G., Mavropoulos, P., Lounis, S. \& Bl\"{u}gel, S. Thermally activated magnetization reversal in monatomic magnetic chains on surfaces studied by classical atomistic spin-dynamics simulations.
\textit{J. Phys.: Condens. Matter} \textbf{23}, 394204 (2011).

\bibitem{Berg81}
Berg, B. \& L\"uscher, M. Definition and statistical distributions of a topological number in the lattice O(3) $\sigma$-model. \textit{Nuclear Physics B} \textbf{190}, 412-424 (1981).

\end{thebibliography}

\begin{thebibliography}{99}

\bibitem{Dzyaloshinskii}
Dzyaloshinskii, I. E. Theory of helicoidal structures in antiferromagnets. III, \textit{Sov. Phys. JETP} \textbf{20}, 665-668 (1965).

\bibitem{Izumov84}
Izyumov, Y. A. Modulated, or long-periodic, magnetic structures of crystals. \textit{Sov. Phys Usp.} \textbf{27}, 845 (1984).

\bibitem{Rybakov_15}
Rybakov, F. N., Borisov, A. B., Bl\"ugel, S. \& Kiselev, N. S. New type of stable particlelike states in chiral magnets. \textit{Phys. Rev. Lett.} \textbf{115} 117201 (2015). 

\bibitem{Bogdanov_89}
Bogdanov, A. N. \& Yablonskii, D. A. Thermodynamically stable "vortices" in magnetically ordered crystals. The mixed state of magnets. \textit{Sov. Phys. JETP} \textbf{68}, 101-103 (1989).

\bibitem{Berg81}
Berg, B. \& L\"uscher, M. Definition and statistical distributions of a topological number in the lattice O(3) $\sigma$-model. \textit{Nuclear Physics B} \textbf{190}, 412-424 (1981).

\end{thebibliography}
\end{document}